\documentstyle[psfig]{article}
\begin{document}
\thispagestyle{empty}
\pagenumbering{arabic}
\title{{\bf Multifractal Behaviour of {\em n}-Simplex Lattice}}
\author{{\bf Sanjay Kumar$^1$\footnote{Present Address: 
Institute for Theoretical Physics,
University of Koln, Zulpicher Street 77 D 50937, Koln, Germany;
email: kumar@thp.Uni-Koeln.DE}, D. Giri$^2$ and Sujata Krishna$^3$} \\
{$^1$\em Department of Physics, Banaras Hindu University, 
Varanasi 221 005, India} \\
{$^2$\em Centre for Theoretical Studies, IIT, Kharagpur 
721 302,  India} \\
{$^3$\em School of Engineering \& Adv. Technology,} \\
{\em Staffordshire University, Stafford ST18 0AD, U.K.}}
\date{} 
\maketitle
\begin{abstract}
We study the asymptotic behaviour of resistance scaling and 
fluctuation of resistance that give rise to flicker noise in 
an {\em n}-simplex lattice.  We propose a simple method to 
calculate the resistance scaling and give a closed-form formula 
to calculate the exponent, $\beta_L$, associated with 
resistance scaling, for any {\em n}. Using current cumulant method 
we calculate the exact noise exponent for {\em n}-simplex lattices.
\end{abstract}

\section{Introduction}
In recent years considerable attention has been devoted to studying 
the properties of disordered systems with the hope of understanding 
percolative phenomena. Key to several such approaches has been the 
concept of randomness and also of frustration [1-12]. However, many 
of the patterns we encounter in nature are not random but self-similar 
and scale invariant [13-14]. For instance, the complicated and scale 
invariant structures that occur when a solid mixture evolves via an 
aggregation process [15]. To understand such systems the concept of 
fractals has been found to be very useful. Fractals are scale 
invariant objects that may be considered as intermediate lattices 
between regular and random (disordered) lattices [13-14,16-17]. Such 
a fractal lattice describes a class of random systems where the 
consequence of the loss of translational invariance of a lattice can 
be studied in detail.  Additionally, resulting from their dialational 
symmetry, statistical, mechanical and transport problems are solvable; 
hence the attraction of the model in such studies [14].

In this paper we consider a particular class of fractal known as the  
{\em n}-simplex lattices, to model various properties of inhomogeneous 
materials [16-17]. The lattice is defined recursively. The map of the 
zero-order truncated {\em n}-simplex lattice is a complete set of 
{\em (n+1)} points. The map of the {\em (r+1)}th order {\em n}-simplex 
lattice is obtained by replacing each of the lattice points of the 
{\em r}th order map by the entire {\em r}th order map. Each of the 
resulting {\em n} points is connected to one of the lines connecting 
the original {\em r}th order vertices. The fractal and spectral dimensions 
of this lattice are given by: 
\begin{equation}
\overline{d} = \frac{\ln(n)}{\ln 2} 
\end{equation}
and 
\begin{equation}
\tilde{d} = \frac{2 \ln (n)}{\ln (n+2)}.
\end{equation}
The lattices with $n \ge 3$ are of particular interest as they provide 
a family of fractals in which $\overline{d}$ varies with $n$ leaving 
$\tilde{d}$ almost constant.

In order to understand how resistance scales with the size of the 
system, in a homogeneous system, we study the distribution of currents 
in a network modeled by an {\em n}-simplex lattice. We consider each 
bond, where bond refers to a line joining two lattice points, of the 
zero-order network as a unit resistor offering resistance $\bf{R}$. 
A unit current enters the network at one of the external nodes and 
leaves through another, the rest of nodes being left open. It is known 
that the distribution of currents in such a network is found to be 
multifractal [18] in the sense that different moments of the distribution 
scale with different exponents.

For resistance scaling analysis two methods may be adopted, either 
(a) to obtain the distribution of current over the entire network and 
measure the energy dissipated in the system or (b) to simplify the network 
and obtain a closed-form solution. This second method has been used 
rigorously by us for resistance scaling and the results obtained by this 
method match those from the current distribution method. The moments of the 
current in a {\em n}-simplex are: 
\begin{equation}
S^{a}_{r}(I_1, I_2,.....I_n) = \sum_{p}|I_{p}|^{a}, 
\end{equation}
where $I_{p}$ is the current in the {\em p}th bond and $p$ goes from 
$1$ to $n(n-1)/2$; $S^{a}_{r}$ is the cumulant for an arbitrary exponent 
$a$. The currents flowing in at the external nodes of a $n$-simplex 
are represented by $I_1$, $I_2$,............$I_n$ respectively 
(see figure 1) with the condition $I_1 + I_2 + ........I_n = 0$.
A scaling factor independent of $I_1$, $I_2$ ........ $I_n$ 
can then be defined as: 
\begin{equation}
\lambda(a) = \frac{S^{a}_{r+1}(I_1,I_2,........I_n)}{S^{a}_{r}
(I_1,I_2,.........I_n)}.
\end{equation}

Note that $\lambda(a)$ is related to the fractal scaling exponent 
$D(a)$. For a fractal with a resistance scaling parameter of 2, 
the {\em r}th generation length scales as $L_{r} = 2^{r}$. Using the 
definitions $S^{a}_{r}(I_1,I_2,.........I_n) = L_{r}^{D(a)}$ and 
$S^{a}_{r}(I_1,I_2,.........I_n) \propto \lambda^{r}(a)$, we get: 
\begin{equation}
D(a) = \frac{\ln \lambda(a)}{\ln 2}.
\end{equation}

The case $a=0$ determines the fractal dimension of the simplex because 
$\lambda(0)$ is simply the ratio of number of bonds in successive order 
of the {\em n}-simplex lattice; $a = 2$ measures the heat loss in the 
network and gives resistance scaling. It has been shown [19-20] that:
\begin{equation}
R(L) \sim L^{-\beta_{l}} \hspace{5.0pt} (L \gg 1)
\end{equation} 
where $\beta_{l}$ is an exponent controlling the transport properties.

In disordered material, the elastic scattering of the carriers at 
impurities leads to the random conductance or resistance fluctuation. 
The fluctuation arises from the interference of the scattered waves, 
and they are random. The magnitude of the resistance noise spectrum 
(flicker noise 1/f) depends on a new exponent, $b$, pertaining to the 
fractal lattice. This exponent (corresponding to $a=4$) is a member 
of infinite number of exponents required to characterize the fractal 
lattice [18]. The exact reason as to why this fluctuation occurs is 
unknown though it is believed that it appears in response to changes 
in many extrinsic parameters such as the carrier density, 
the applied measuring current, external electric fields and external 
magnetic fields. The spectrum of resistance fluctuation is given by 
\begin{equation}
S_{R}(w) = \int e^{iwt} <R(t)R(0)> dt.
\end{equation}

The exponent $b$ associated with the scaling behaviour of normalized 
noise is given by [19]: 
\begin{equation}
\rho_R = \S_{R}/R^2 \sim L^{-b} \hspace{5.0pt} (L >> 1)
\end{equation}
As long as each bond resistance fluctuates independently with the same spectrum,
the explicit frequency dependence can be discarded. The upper and lower bounds of $b$ [19-20] are given by: 
\begin{equation}
\beta_{L} < b < \overline{d}.
\end{equation}

The paper is organised as follows: In Section 2 we derive a closed-form 
solution to calculate $\beta_{L}$ for any value of {\em n}. In Section 3 
we use the current cumulant method to calculate the noise exponent for 
{\em n}-simplex. The paper ends with a brief discussion on the 
bounds proposed and comparison with our results with experimental data.

\section{Calculation of $\beta_L$ associated with resistance scaling}

In this section we propose a simple method of calculating $\beta_{l}$ 
for any {\em n}-simplex. Consider a fixed current $I_1$ entering at one 
of the external nodes of the lattice, and leaving from another, all the 
remaining external nodes being left open ($I_2 = I_3 =...I_n = 0$).
We calculate the equivalent 
resistance between these two external nodes and establish a recursion 
relation between {\em r}th and {\em (r+1)}th order lattices and use the 
Real Space Renormalization Group Technique to find the exponents 
[18,21-22].

>From the symmetry properties of the simplex, it is apparent that all 
{\em (n-2)} external nodes apart from those through which current 
enters and leaves are at equipotential. Redrawing just those bonds 
through which currents flow, we have {\em (n-1)} parallel paths for 
current to flow.  Of these paths, one offers unit resistance and each 
of the others offer twice the unit resistance (since they include two 
resistances in series). Hence the equivalent resistance is given by: 
\begin{equation}
\frac{1}{R_{E}} = \frac{1}{R} + [\frac{1}{2R} + \frac{1}{2R} + 
\cdots  (n-2) \; {\rm terms}] 
\end{equation}
where R is the unit resistance and the square bracket contains exactly 
(n-2) identical terms. This directly leads to: 
\begin{equation}
R_{E} = \frac{2R}{n}.
\end{equation}

Now if we consider a star of {\em n}-branches, each offering a 
resistance of $R/n$, the effective resistance between any two external 
nodes through which current flows will be $2R/n$ as they are in series 
and all other nodes being left open. It is then straight forward to show 
using these transformation for $n$-simplex lattice that the following 
scaling holds good:
\begin{equation}
\lambda(2) \sim \frac{R(2L)}{R(L)} =\frac{R_{r+1}}{R_{r}} =  
\frac{n+2}{n}. 
\end{equation}

>From equation (11) we know that for any {\em n}-simplex the equivalent 
resistance of first order is: 
\begin{equation}
R_{E1} = \frac{2R}{n},
\end{equation}
combined with equation (12) gives the  equivalent resistance of 
{\em r}th order as: 
\begin{equation}
R_{Er} = \frac{2(n+2)^{r-1}}{n^r}.
\end{equation}

Thus we see how, by merely knowing the simplex one can calculate the 
equivalent resistance of any iteration. No long winded applications of 
Kirchoff's Laws are required to obtain the resistance scaling. The 
exponent $\beta_{L}$ is related to $\lambda(2)$ by: 
\begin{equation}
\beta_{L} = \frac{\ln (1/ \lambda(2))}{\ln 2}.
\end{equation}

\section{Calculation of the Flicker noise exponent on $n$-simplex}

It has been shown that  the $4^{th}$  moment of current distribution 
is associated with the noise exponent [18].

Assuming that the cumulant $S^{4}_{r}(I_1, I_2,.....I_n)$ can be 
expressed as a homogeneous polynomial of degree $4$, the most 
general polynomial is a linear combination of $P_1^4$, $P_1^2 P_2$,
$P_1 P_3$, $P_4$ and $P_2^2$. These polynomials are defined as
\begin{eqnarray*}
P_1 & = & I_1 + I_2 + I_3 + ............I_n  \\
P_2 & = & I_1^2 + I_2^2 + I_3^2 + ............I_n^2 \\
P_3 & = & I_1^3 + I_2^3 + I_3^3 + ............I_n^3 \; {\rm and} \\
P_4 & = & I_1^4 + I_2^4 + I_3^4 + ............I_n^4  \\
\end{eqnarray*}

However in present case $P_1 = 0$ due to current conservation. 
Hence $S^{4}_{r}(I_1, I_2,.....I_n)$ can be written as 
\begin{equation}
S^{4}_{r}(I_1, I_2,.....I_n) = A_r P_2^2 (I_1, I_2,.....I_n) +
B_r P_4 (I_1, I_2,.....I_n)
\end{equation}

The next step is to determine $S^{4}_{r-1}(I_1, I_2,.........I_n)$. 
To establish a recursion relation between \\ 
$S^{4}_{r} (I_1, I_2,.......I_n)$ and $S^{4}_{r-1}(I_1, I_2,......I_n)$ 
we obtained current distribution in an $n$-simplex lattice at each node. 
It is easy to see that the current distribution at each node is
\begin{displaymath}
\sum_{k=1}^{n} \left [I_k + \frac{1}{n-1} \sum_{j=1 \atop k\ne j}^{n} 
I_k - I_j\right]
\end{displaymath}
by current conservation. In figure 1 we have shown the current along 
each bond. Therefore, we can write
\begin{displaymath}
S^{4}_{r}(I_1, I_2,.....I_n) = S^{4}_{r-1}(I) + S^{4}_{r-1}(II) +
..............+ S^{4}_{r-1}(n)
\end{displaymath}
where $S_{r-1}(I)$, $S_{r-1}(II),........$ are the current cumulants of
$(r-1)$th order of $n$-simplex lattice. $I$, $II,......$ represents 
shaded region in figure 1. Above equation can be expressed as 
\begin{eqnarray}
S^{4}_{r}(I_1, I_2,.....I_n) & = & A_{r-1} P_2^2 (I_1, I_2,.....I_n) +
B_{r-1} P_4 (I_1, I_2,.....I_n) \nonumber \\ 
& + & A_{r-1} P_4 (I_1, I_2,.....I_n) + B_{r-1} P_2^2 (I_1, I_2,.....I_n)
\end{eqnarray}
which establish the transformation relation between $r$ and $(r-1)$th
order. Comparing equations (16) and (17) we obtain the recursion relation 
between the $n$-simplex lattices of the $r$ and $(r-1)$ th order.
\begin{equation}
\pmatrix{A_r \cr B_r \cr}  = \frac{1}{n^4} \pmatrix{(n^3 +2)n & 
n^2 (n+1)^2 \cr 6 & (2n + 3)n \cr} 
\pmatrix{A_{r-1} \cr B_{r-1} \cr}   
\end{equation}
The eigenvalues corresponding to the transformation matrix for the
$n$-simplex lattice are given by
\begin{displaymath}
\lambda_n^{\pm} (a = 4) = \frac{(n^3 + 2 n + 5)n \pm n(n+1)
\sqrt{n^4 - 2 n^3 - n^2 + 2 n +25}}{2 n^4}
\end{displaymath}
and the fractal scaling exponent corresponding to largest eigenvalue is 
\begin{displaymath}
D(a=4) = \frac{\ln\lambda_n^{+} (a=4)}{\ln 2}.
\end{displaymath}

\section{Discussion}
We have seen that various moments of branch current give rise to 
different exponents, namely exponent $b$ associated with the noise 
amplitude, $\beta_{L}$ associated with resistance scaling and 
$\overline{d}$ associated with the mass of the fractal. The relation 
between $D(4)$ obtained in Section 3 and the exponent $b$ for 
normalised noise is as follows:
\begin{equation}
\frac{S_{R}}{R^{2}_{r}} \sim \rho_{R} \sim L_{r}^{-b}.
\end{equation}

Now, $S_{R_{r}} \sim \sum_{p} |I_{p}|^{4}$
\begin{equation}
\frac{\sum_{p} (I_{p})^{4}_{r}}{R^{2}_{r}} \sim \rho_{R_{r}} \sim 
L_{r}^{-b}.
\end{equation}

This gives:
\begin{equation}
\frac{\rho_{R_{r+1}}}{\rho_{R_{r}}} = \lambda (\rho_{R_{r}}) \sim 
2^{-b}
\end{equation}
or,
\begin{equation}
b = \frac{\ln (1/ \lambda (\rho_{R}))}{\ln 2}.
\end{equation}

Now,
\begin{equation}
\lambda(\rho_{R}) = \frac{\sum_{p} (I_{p})^{4}_{r+1}}{\sum_{p} 
(I_{p})^{4}_{r}} \frac{R_{r}^{2}}{R_{r+1}^{2}}
\end{equation}
or,
\begin{equation}
\lambda(\rho_{R}) = \frac{ A_{r+1}}{A_{r}} \left( \frac{R_{r}}{R_{r+1}} 
\right)^{2}
\end{equation}
\begin{equation}
\lambda(\rho_{R}) = \frac{ A_{r+1}}{A_{r}} \left( \frac{1}{\lambda(R)} 
\right)^{2}.
\end{equation}

Substituting this in equation(8) gives: 
\begin{equation}
b = \frac{\ln \left(\lambda^{2}(R) \times 
\frac{A_r}{A_{r+1}}\right)}{\ln 2}.
\end{equation}

This expression gives the respective values of the exponent $b$ as 
$1.1844$, $1.0629$ and $0.9269$ for the 3, 4 and 5-simplex respectively. 
It is clear that the inequality $\overline{d} \ge b \ge -\beta_{L}$ 
is satisfied in each of the three cases.

In the limit $n$ goes to infinity $\lambda(2) = (n+2)/n$ goes to 1. 
This is due to the fact that a large number of parallel equi-resistance 
paths are available for current flow. Such a large number of paths are 
available that in going from one order to the next we are in effect not 
altering the equivalent resistance. The exponent $\beta_{L}$ decreases 
in magnitude as we go to higher dimension, implying that resistance becomes 
less dependent on the length of the fractal.

With regard to flicker noise, we have seen that the scaling relation 
becomes increasingly complex as the order of simplex is increased.
The noise versus resistance exponent $Q$ is defined by the following: 
\begin{equation}
Q = 2 + \frac{t}{k}
\end{equation}
where $t$ and $k$ are given by:
\begin{equation}
R \sim (\Delta p)^{-t}
\end{equation}
and
\begin{equation}
\frac{S_{R}}{R^2} \sim (\Delta p)^{-k}.
\end{equation}

The experimental measurements [19-20] of $t$ and $k$ were made on 
2d-carbon-vax mixtures and found to be $2.3 \pm 0.4$ and $5 \pm 1$ 
respectively.  The direct plot of $S_{R}$ versus $R$ leads to $S_{r} 
\sim R^{Q}$ where $Q = 3.7 \pm 0.2$. The value we obtain for $Q$ is 
in agreement with this as is clear from Table 1.

However, similar measurements on two dimensional films and metallic 
films have given values of $Q$ differing from what we predict. 
Perhaps instead of taking the $n$-simplex lattice, if one considered 
a 2-d Sierpinski gasket [13-14] better results could be expected.  
For all $n > a$, there is a finite dimension matrix whose largest 
eigenvalue will give the characteristic exponents. The matrix elements 
are function of $n$ and hence eigenvalues will be the well defined 
function of $n$. But for $n < a$ the result will be obtained by smaller 
matrix. The generalization to higher value of $a$ and rescaling factor 
$b > 2$ is under  progress.

\section*{Acknowledgements}
\noindent
We would like to thank Yashwant Singh and Deepak Dhar for many helpful 
discussions. Financial assistance from the Department of Science and 
Technology India is acknowledged.  One of us(SK) would like to thank 
INSA-DFG for financial support.

\newpage
\section*{References}
\begin{enumerate}
\item S. Washburn and R. A. Webb,{\em Reports on Progress in Physics}, 
$\bf 55$, 1311 (1992) 
\item A. K. Sen, {\em Modern Physics Letter B}, $\bf 11$, 555 (1997)
\item V. I. Kozub and A. M. Rudin, {\em Phy. Rev. B}, $\bf 53$, 5356 
(1996)
\item A. G. Hunt, {\em J. Phys: Condensed Matter}, $\bf 10$, L303 
(1998)
\item A. K. Gupta, A. M. Jayannavar and A. K. Sen, {\em J. Phys.(Paris)}, 
$\bf 3$, 1671 (1993)
\item M. James, et.al, {\em Phy. Rev. Lett.}, $\bf 56$, 2280 (1986)
\item Y. Gefen, A. Aharony, B. B. Mandelbrot and S. Kirkpatrick, {\em 
Phy.  Rev. Lett.}, $\bf 47$, 1771, (1981)
\item B. W. Southern and A. R. Douchant, {\em Phy. Rev. Lett.}, $\bf 
55$, 1148 (1985)
\item B. Docut and R. Rammal, {\em Phy. Rev. Lett.}, $\bf 55$, 1148, 
(1985)
\item I. Zivic, S. Milosevic, and H. E. Stanley, {\em Phy. Rev. E}, 
$\bf 47$, 2340 (1990)
\item D. C. Hong, and H. E. Stanley, {\em J. Phys. A}, $\bf 16$, L525 
(1983)
\item H. J. Herrmann, D. C. Honmg and H.E. Stanley, {\em J. Phys. A}, 
$\bf 17$, L261 (1984)
\item B. B. Mandelbrot, {\em The Fractal Geometry of Nature}, (Freeman, 
NY 1982)
\item L. Pietronero and E. Tosatti (eds) {\em Fractals in Physics}, 
(North Holland: Amsterdam 1986)
\item D. Kessler, J. Koplick and H. Levine, {\em Adv. in Phys.}, $\bf 
37$, 255 (1988)
\item D. R. Nelson and M. E. Fisher, {\em Ann. Phys.}, $\bf 91$, 266 
(1975)
\item D. Dhar, {\em J. Math. Phys.}, $\bf 18$, 577 (1977)
\item S. Roux and C. D.  Mitescu, {\em Phy. Rev. B }, $\bf 35$, 898 
(1987)
\item R. Rammal, C. Tannous and A.M.S. Tremblay, {\em Phy. Rev. Lett.}, 
$\bf 54$, 1718 (1985)
\item R. Rammal C. Tannous and A. M. S. Tremblay, {\em Phy. Rev. A}, 
$\bf 31$, 2662 (1985)
\item P. Alstrom, D. Stassinopoulos and H. E. Stanley, {\em Physica A}, 
$\bf 153 $, 20 (1988)
\item P. Y. Tong and K. W. Yu, {\em Phys. Lett A}, $\bf 160$, 293 
(1991)
\end{enumerate}

\newpage

\begin{center}
\psfig{figure=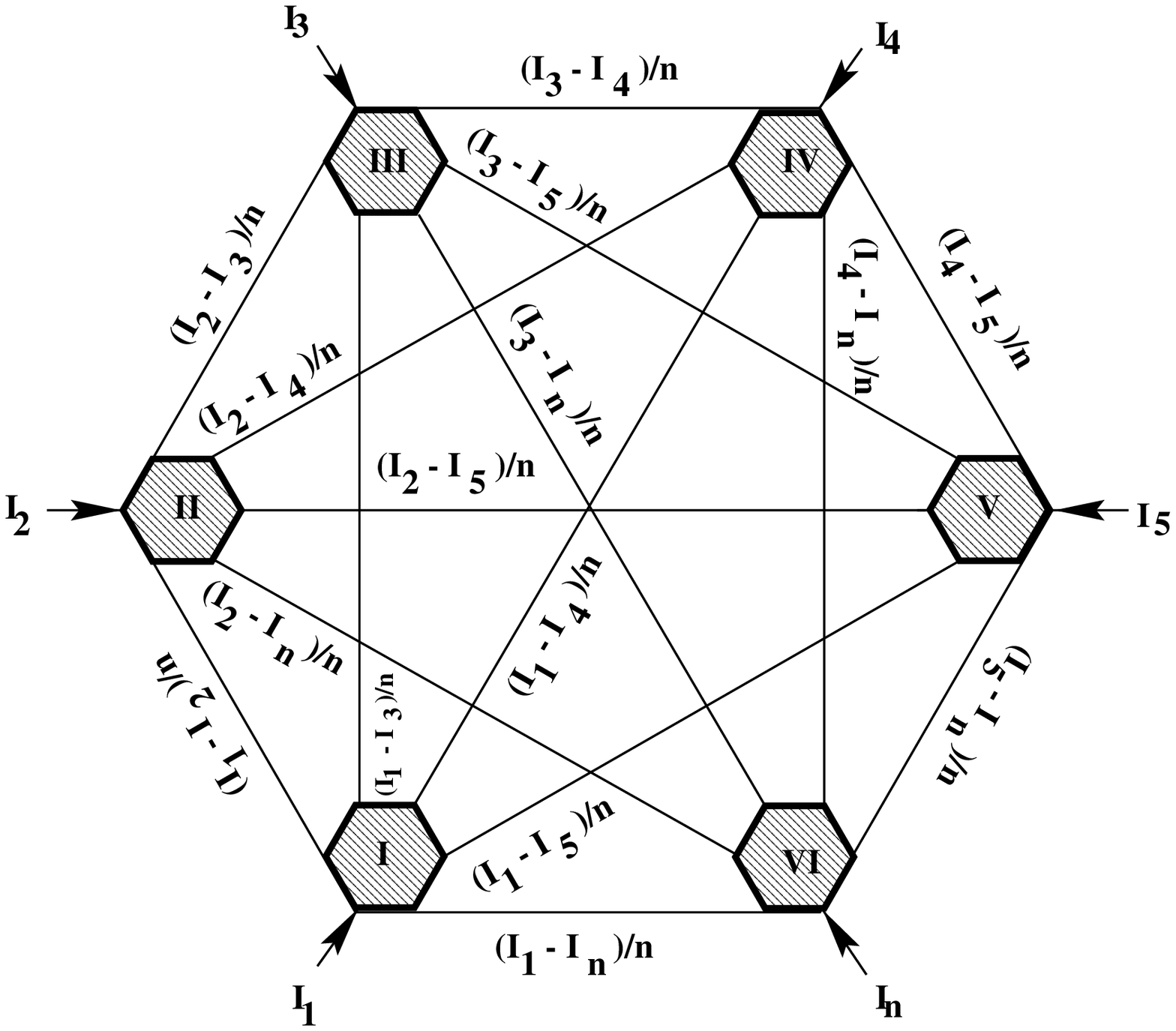,height=5in}
\end{center}

{\bf Figure 1 :} Schematic representation of $n$-simplex lattice 
($n=6$). The current along each bond has been shown.

\newpage

\begin{table}
\caption{{\it Exponents calculated for $3$-, $4$- and $5$-simplex 
lattices.}}
\begin{center}
\begin{tabular}{|l|l|l|l|l|} \hline
& & & & \\
simplex & $\bar{d}$ & b & $\beta_{L}$ & Q \\ 
& & & & \\ \hline
3 & 1.585 & 1.184 & 0.737 & 3.607 \\
& & & & \\
4 & 2.000 & 1.063 & 0.585 & 3.817 \\
& & & & \\
5 & 3.322 & 0.927 & 0.485 & 3.909 \\ \hline
\end{tabular}
\end{center}
\end{table}

\end{document}